\pdfoutput=1
\def\Title{Riemann Normal Coordinate expansions using Cadabra.}
\def\Author{Leo Brewin}
\documentclass[a4paper,12pt]{article}
\usepackage[style=numeric, sorting=none, sortcites=true]{biblatex}
\usepackage{cdblatex}
\usepackage{hyperref}
\usepackage{geometry}
\usepackage{brewin}

\hypersetup{colorlinks=true,
            citecolor=blue,
            linkcolor=red,
            pdfauthor= \Author,
            pdftitle = \Title}

\numberwithin{equation}{section}

\begin{document}

\input{./cadabra/export.cdbtex}

\CdbSetup{action=verbatim}

\lstset{numbers=left,gobble=2,basicstyle=\ttfamily\small,escapechar=|}

\title{\Title}
\author{\Author\\[10pt]%
School of Mathematical Sciences\\%
Monash University, 3800\\%
Australia}
\date{12-Aug-2009 (v1)\\
      10-Dec-2019 (v2)\\
       4-Nov-2022 (v3)}
\reference{Preprint: arXiv:0903.2087\\[5pt]
           Journal: {\it Class.Quantum.Grav.} {\bf 26} 175017 (2009)}

\maketitle

\begin{abstract}
\noindent
Riemann normal coordinate expansions of the metric and other geometrical quantities,
including the geodesic arc-length, will be presented. All of the results are given to
fifth-order in the curvature and were obtained using the computer algebra package Cadabra.

\end{abstract}

\BegNote{\bf\large Version 2}
The first version of this paper was written in 2009. This second version (written in 2019)
differs from the first in a number of important ways. First, the discussion on Cadabra has
been completely rewritten to account for the revised syntax introduced in the latest version
of Cadabra (itself updated in 2016)\footnote{The Cadabra results in this paper are based on
\CdbVersion}. Second, and more importantly, a significant computational error in the first
version (of this paper) has been corrected. Though the algorithms given in the first version
were correct their implementation in Cadabra (by this author) was incorrect. This lead to
errors first in the quartic and higher order terms in the generalised connections and then,
down the line, to computations built upon the generalised connections (e.g., the initial and
boundary value problem and the arc-length equations). The Cadabra codes have been corrected
and thoroughly cross-checked. A number of minor typographical errors have also been
corrected.

\BegNote{\bf\large Version 3}
This version corrects the indicies (and sign) of the left hand side of equations 11.19a-d.
This brings the arXiv version in line with the version on GitHub
\ \url{https://github.com/leo-brewin/riemann-normal-coords}.
Many thanks to David Lowe for spotting these typos.
\EndNote

\section{Introduction}

In a previous paper \cite{brewin:2009-02} a series of simple examples were used to
demonstrate how the computer algebra program Cadabra
(\cite{peeters:2007-02},
\cite{peeters:2018-01},
\cite{peeters:2017-01},
\cite{peeters:2017-02})
could be employed to do the kinds of tensor computations often encountered in General
Relativity. The examples were deliberately chosen to be sufficiently simple as to allow the
reader to appreciate the basic functionality of Cadabra. This left open the question of how
well Cadabra might perform on more challenging computations. The purpose of this paper is to
address that point by showing how Cadabra can be used to express various geometrical
objects, such as the metric, the connection and the geodesic arc-length, in terms of Riemann
normal coordinates (also known as geodesic coordinates). This is a standard computation in
differential geometry that leads to a series expansion for the various geometrical objects
in powers of the curvature and its derivatives. The first few terms are rather easy to
compute by hand but further progress, to higher order terms, quickly becomes prohibitively
difficult. Thus some computer tool, in this case Cadabra, is essential when computing these
higher order terms.

The first part of this paper will be concerned with purely mathematical and algorithmic
issues including a definition of Riemann normal coordinates, how they can be constructed
(from generic coordinates) and what properties they confer upon geometrical objects (such as
the metric, the connection etc.). In the second part the attention will shift to issues
specific to Cadabra. This will be followed by full details of the 5th-order expansion of the
metric and other objects.

It should be emphasised that the primary objective of this paper is to demonstrate that
Cadabra is a mature program that can easily handle an otherwise prohibitive computation.
Some of the Cadabra results for the metric do replicate results by other authors (e.g.,
equation \eqref{eqn:metric} is consistent with that given in
\cite{hatzinikitas:2000-01},
\cite{higashijima:2002-01},
\cite{yamashita:1984-01},
and
\cite{muller:1997-01}
while some results are original (in particular, the solution for the two-point boundary value
problem, \eqref{eqn:bvp6}).

Before turning the discussion to the definition and construction of Riemann normal
coordinates, a simple coordinate transformation will be introduced. This will help in making
clear what is meant by the statement that an expression is an expansion to a given order.

\section{Conformal coordinates}\label{sec:Conformal}

Each algorithm given later in this paper yields polynomial approximations to particular
geometric quantities (e.g., the metric). Higher order approximations are obtained by
recursive application of the algorithms.

The primary goal of this section is to provide meaning to the statement that the polynomial
$S_{\eps}$ is an expansion of $S$ up to and including terms of order $\BigO{\eps^n}$. The
key to this definition is the use of a conformal transformation of the original metric.

Consider some neighbourhood of $O$ and let $\eps$ be a typical length scale for $O$ (for
example, $\eps$ might be the length of the longest geodesic that passes through $O$ and
confined by the neighbourhood). Construct any regular set of coordinates $x^a$ (i.e., such
that the metric components are non-singular) in the neighbourhood of $O$ and let the
coordinates of $O$ be $x^a_{\star}$. The word \emph{patch} will be used to denote the
neighbourhood of $O$ in which these coordinates are defined. Now define a new set of
coordinates $y^a$ by
\[
x^a = x^a_{\star} + \eps y^a
\]
and thus
\[
  ds^2 = g_{ab}(x)dx^a dx^b
       = \eps^2 g_{ab}(x_{\star} + \eps y) dy^a dy^b
\]
The conformal metric $d\tilde s$ can now be defined by $d\tilde s = ds/\eps^2$. This leads to
\[
d{\tilde s}^2 = g_{ab}(x_{\star} + \eps y) dy^a dy^b
              = {\tilde g}_{ab}(y,\eps) dy^a dy^b
\]
and
\[
{\tilde g}_{ab} = g_{ab}\>,\qquad
{\tilde g}_{ab,c} = \eps g_{ab,c}\>,\qquad
{\tilde g}_{ab,cd} = \eps^2 g_{ab,cd}\qquad \text{at }O
\]
where the partial derivatives on the left are with respect to $y$ and those on
the right are with respect to $x$. Since $g_{ab}(x_{\star})$ does not depend on $\eps$
it is easy to see that
\begin{equation*}
{\tilde g}_{ab,i_1i_2i_3\cdots i_n} = \BigO{\eps^n}\qquad \text{at }O
\end{equation*}
From this it follows, by simple inspection of the standard equations, that
\begin{align*}
{\tilde\Gamma}^{a}{}_{bc,i_1i_2i_3\cdots i_n} &= \BigO{\eps^{n+1}}\qquad \text{at }O\\[5pt]
{\tilde R}^{a}{}_{bcd,i_1i_2i_3\cdots i_n} &= \BigO{\eps^{n+2}}\qquad \text{at }O
\end{align*}

There are now two ways to look at the patch. It can be viewed as a patch of length scale
$\eps$ with a curvature independent of $\eps$. Or it can be viewed as a patch of fixed size
but with a curvature that depends on $\eps$ (and where the limit $\eps\rightarrow0$
corresponds to flat space). This later view is useful since it ensures that the series
expansions around flat space can be made convergent (by choosing a sufficiently small
$\eps$).

These conformal coordinates will be used for the remainder of this paper. As there is no
longer any reason to distinguish between $x^a$ and $y^a$, the $y^a$ will be replaced with
$x^a$. The $x^a$ will now be treated as generic coordinates (but keep in mind that there is
an underlying conformal transformation built into these coordinates).

Finally, the statement that $S_\eps$ is an order $\BigO{\eps^n}$ expansion of $S$ will be
taken to mean that $n$ is the smallest positive integer for which
\[
0 < \lim_{\eps\rightarrow0} \frac{\vert S-S_\eps\vert}{\eps^{n+1}} < M
\]
for some finite positive $M$ i.e., if $S$ were expanded as a Taylor series in $\eps$ around
$\eps=0$ then $S$ and $S_\eps$ would differ by terms proportional to $\eps^{n+1}$.



\section{Riemann Normal Coordinates}\label{sec:RNC}

The basic idea behind Riemann normal coordinates is to use the geodesics through a given
point to define the coordinates for nearby points. Let the given point be $O$ (this will be
the origin of the Riemann normal frame) and consider some nearby point $P$. If $P$ is
sufficiently close to $O$ then there exists a unique geodesic joining $O$ to $P$. Let $v^a$
be the components of the unit tangent vector to this geodesic at $O$ and let $s$ be the
geodesic arc length measured from $O$ to $P$. Then the Riemann normal coordinates $x^a$ of
$P$ relative to $O$ are defined to be $x^a = s v^a$. These coordinates are well defined
provided the geodesics do not cross (which can always be ensured by choosing the
neighbourhood of $O$ to be sufficiently small). For more mathematical details see
(\cite{chern-chen-lam:2000-01},
\cite{chavel:2006-01},
\cite{eisenhart:1926-01})
and in particular the elegant exposition by Gray
(\cite{gray:1973-01}, \cite{willmore:1996-01}).

One trivial consequence of this definition is that all geodesics through $O$ are of the form
$x^a(s) = s v^a$ and that the $v^a$ are constant along each geodesic. This implies, by
direct substitution into the geodesic equation, that $\Gamma^c{}_{ab}=0$ at $O$ which in
turn implies that $g_{ab,c}=0$ at $O$. Suppose now that the metric can be written as a
Taylor series in $x^a$ about $O$. In that series only the zero, second and higher
derivatives of the $g_{ab}$ would appear. Thus the leading terms of the metric can be
expressed as a sum of a constant part (the leading term) plus a curvature part (from the
second and higher derivative terms). Similar expansions arise for other geometrical
quantities (e.g., geodesics, arc length) in terms of a flat space part plus a curvature
contribution.

Suppose, as is almost always the case, that the coordinates $x^a$ are \emph{not} in Riemann
normal form. How might they be transformed to a local set of Riemann normal coordinates? A
first attempt might be to make direct appeal to the basic definition, namely, $y^a = s v^a$.
Unfortunately this does raise an immediate problem. The quantities $v^a$ are rarely known
explicitly but must instead be computed by solving a two-point boundary value problem. This
is non-trivial but it can be dealt with in a number of ways. The approach adopted here is
to iterate over a sequence of initial value problems to provide a sequence of approximations
to the solution of the original boundary value problem. Here is a brief outline of the
process -- this serves as an introduction to the next two sections where the full details
will be given.

First, consider any method for computing the coordinates $x^a(s)$ of a typical geodesic that
originates from $O$. This solution of the typical initial value problem will depend on two
integration constants, $x^a$ and ${\Dot x}^a$, being the respective values of $x^a(s)$ and
$dx^a/ds$ at $s=0$. Next, embed this first step in an iterative scheme (e.g., a fixed-point
scheme) to produce successive approximations to ${\Dot x}^a$ so that the geodesic passes
through not only $O$ but also $P$. The final step, assuming the sequence converges, is to set
$v^a$ to be ${\Dot x}^a$ and finally $y^a = s v^a$.

The next two sub-sections provide further details on how the initial and boundary value
problems were solved.

\subsection{The initial value problem}\label{sub:CoordsIVP}

The aim here is to obtain a Taylor series, about the point $O$, for the solution of the
geodesic equation
\[
0 = \frac{d^2x^a}{ds^2} + \Gamma^a{}_{bc}(x) \frac{dx^b}{ds}\frac{dx^c}{ds}
\]
subject to the initial conditions $x^a(s) = x^a$ and $dx^a/ds={\Dot x}^a$ at $s=0$.

First choose $s=0$ at $O$ and then write the Taylor series for $x^a(s)$ as
\[
x^a(s) = \left.x^a\right\vert_{s=0}
       + s \left.\frac{dx^a}{ds}\right\vert_{s=0}
       + \sum_{n=2}^\infty\> \frac{s^n}{n!}\left.\frac{d^nx^a}{ds^n}\right\vert_{s=0}
\]
The second and higher derivatives can be obtained by successive differentiation of the
geodesic equation leading to
\begin{equation}
x^a(s) = x^a + s {\Dot x}^a
       - \sum_{n=2}^\infty \frac{s^n}{n!}\Gamma^{a}{}_{i_1i_2i_3\cdots i_n}
                                         {\Dot x}^{i_1}
                                         {\Dot x}^{i_2}
                                         {\Dot x}^{i_3}\cdots
                                         {\Dot x}^{i_n}
\pglabel{eqn:SolIVP}
\end{equation}
where the $\Gamma^{a}{}_{i_1i_2i_3\cdots i_n}$, known as \emph{generalised connections},
are defined recursively by
\begin{equation}
\Gamma^{a}{}_{i_1i_2i_3\cdots i_n} = \Gamma^{a}{}_{(i_1i_2i_3\cdots i_{n-1},i_n)}
                                 - n \Gamma^{a}{}_{p(i_2i_3\cdots i_{n-1}}
                                     \Gamma^{p}{}_{i_1i_n)}
\pglabel{eqn:GenGam}
\end{equation}
Note that the use of round brackets $(...)$ denotes total symmetrisation over the included
indices (see Appendix A for more details).

A convenient shorthand for equation \eqref{eqn:GenGam} in terms of covariant derivatives can
be obtained by ignoring (in this context alone) the single upper index. This leads to the
compact notation
\begin{equation}
\Gamma^{a}{}_{i_1i_2i_3\cdots i_n} = \Gamma^{a}{}_{(i_1i_2;i_3i_4i_5\cdots i_n)}
\pglabel{eqn:gamCovDeriv}
\end{equation}

\subsection{The boundary value problem}\label{sub:CoordsBVP}

The question here is: How can ${\Dot x}^a$ be chosen so that the geodesic passes through not
only $O$ but also $P$? The geodesic, by construction, already passes through $O$
so that leaves $P$ as the main focus of attention. Suppose that the coordinates of $P$ are
$x^a + \Delta x^a$ where $x^a$ are the coordinates of $O$. Let $s_P$ be the geodesic
distance from $O$ to $P$. The challenge now is to solve
\[
\Delta x^a = s_P {\Dot x}^a
           - \sum_{n=2}^\infty \frac{s_P^n}{n!}\Gamma^{a}{}_{i_1i_2i_3\cdots i_n}
                                               {\Dot x}^{i_1}
                                               {\Dot x}^{i_2}
                                               {\Dot x}^{i_3}\cdots
                                               {\Dot x}^{i_n}
\]
for ${\Dot x}^a$ in terms of $\Delta x^a$ and the generalised connections.

Put $y^a = s_P {\Dot x}^a$ (this introduces the Riemann normal coordinates) and re-arrange
the equation into the form
\begin{equation}
y^a = \Delta x^a + \sum_{n=2}^\infty \frac{1}{n!}\Gamma^{a}{}_{i_1i_2i_3\cdots i_n}
                   y^{i_1} y^{i_2} y^{i_3}\cdots y^{i_n}
\pglabel{eqn:solveBVP}
\end{equation}
In a small variation to the plan given above, the plan will be modified slightly to solve
this equation for $y^a$ (rather than ${\Dot x}^a$) by constructing a sequence of
approximations $y^a_m$ to $y^a$. The $y^a$ will, through the generalised connections, depend
on the Riemann curvature and its derivatives (at $O$). So in principle the $y^a$ could be
expanded as power series in $\eps$ (i.e., as a power series in the curvatures). The $y^a_m$
will be chosen to be the Taylor polynomial of $y^a$ to order $\eps^m$. That is, $y^a_m$ is a
polynomial in the curvatures (and its derivatives) up to and including terms of order
$\BigO{\eps^m}$. The $y^a_m$ can computed by truncating both sides of \eqref{eqn:solveBVP}
to terms no higher than $\BigO{\eps^m}$. But note that the $\Gamma^{a}{}_{i_1i_2i_3\cdots
i_n}$ are of order $\BigO{\eps^{n-1}}$. The upshot is that the infinite series may be
truncated at $n=m$ while still retaining all terms up to and including $\eps^m$. This leads
to
\[
y^a_m = \Delta x^a
      + T^m_{\eps}\left(
             \sum_{n=2}^{n=m} \frac{1}{n!}\Gamma^{a}{}_{i_1i_2i_3\cdots i_n}
                  y^{i_1}_m y^{i_2}_m y^{i_3}_m\cdots y^{i_n}_m
                  \right)
\]
where $T^m_{\eps}$ is a simple truncation operator (it deletes all terms of order
$\BigO{\eps^{m+1}}$ or higher). This is a marginal improvement on \eqref{eqn:solveBVP} (at
least it is a finite series) but it is still a non-linear equation for $y^a_m$. But
fortunately improvements are readily made. Notice, once again, that
$\Gamma^{a}{}_{i_1i_2i_3\cdots i_n} = \BigO{\eps^{n-1}}$ and this allows the use of lower
order estimates for $y^a$ in the product terms on the right hand side. This leads to
\begin{equation}
y^a_m = \Delta x^a
      + T^m_{\eps}\left(
             \sum_{n=2}^{n=m} \frac{1}{n!}\Gamma^{a}{}_{i_1i_2i_3\cdots i_n}
                  y^{i_1}_{m-n+1} y^{i_2}_{m-n+1} y^{i_3}_{m-n+1}\cdots y^{i_n}_{m-n+1}
                  \right)
\pglabel{eqn:iterBVP}
\end{equation}
This shows that $y^a_m$ appears only on the left hand side and thus this equation can be
used to recursively compute $y^a_p$ for $p=2,3,4,\cdots$. Here are the first few $y^a_m$.
Starting with the lowest order approximation,
\[
y^a_0 = \Delta x^a
\]
and as there are no $\eps^1$ terms in \eqref{eqn:solveBVP} this can also be written as
\[
y^a_1 = y^a_0 = \Delta x^a
\]
Now set $m=2$ in equation \eqref{eqn:iterBVP} to obtain
\begin{align*}
y^a_2 &= \Delta x^a
       + T^2_\eps\left(\frac{1}{2!} \Gamma^a{}_{i_1i_2} y^{i_1}_1 y^{i_2}_1\right)\\[3pt]
      &= \Delta x^a + \frac{1}{2} \Gamma^a{}_{i_1i_2} \Delta x^{i_1}_1 \Delta x^{i_2}_1
\end{align*}
and once more, with $m=3$, with the result
\begin{align*}
y^a_3 &= \Delta x^a
       + T^3_\eps\left(
             \frac{1}{2!} \Gamma^a{}_{i_1i_2} y^{i_1}_2 y^{i_2}_2
            + \frac{1}{3!} \Gamma^a{}_{i_1i_2i_3} y^{i_1}_1 y^{i_2}_1 y^{i_3}_1\right)\\[3pt]
      &= \Delta x^a + \frac{1}{2} \Gamma^a{}_{i_1i_2} \Delta x^{i_1} \Delta x^{i_2}
                    + \frac{1}{6} \left(  \Gamma^a{}_{bi_1}\Gamma^b{}_{i_2i_3}
                                        + \Gamma^a{}_{i_1i_2,i_3}\right) \Delta x^{i_1}
                                                                         \Delta x^{i_2}
                                                                         \Delta x^{i_3}
\end{align*}
This process may seem simple but looks can be deceiving -- the higher order $y^a_m$ contain
a profusion of terms that, when computed by hand, are largely unmanageable beyond $m\approx
7$. At this point there is little option but, if higher order terms are needed, to defer the
computations to a suitable computer algebra package. This point will be discussed in more
detail in section (\ref{sec:Cadabra}).

This completes the first objective -- to find a way to transform from any non-singular set
of coordinates to a local set of Riemann normal coordinates. The question to ask now is --
What form does the metric (and other objects) take in these coordinates? This is the subject
of the second next section. But before doing so, here is a short digression to introduce
some new notation.

\section{Notation}\label{sec:Notation}

It has already been noted that the recursive nature of equations like \eqref{eqn:GenGam}
imposes a significant computational cost with each iteration. A less obvious side effect is
that number of tensor indices on each term will also increase with each iteration. The
equation for the generalised connections, \eqref{eqn:GenGam}, is a case in point. This
contains an index list of the form $i_1i_2i_3\cdots i_n$. Lists such as these are tedious to
write and are prone to transcription error (when in the hands of humans). It thus makes sense
to use a notation that is easy to read and write while also not detracting from the meaning
in the expression. The proposal is that a sequence of indices such as $i_1i_2i_3\cdots i_n$
be replaced with a single index of the form $\ui$. In this notation the equation for the
generalised connections \eqref{eqn:GenGam} would be written as
\[
\Gamma^{a}{}_{b\uc d} = \Gamma^{a}{}_{(b\uc,d)}
                - (n+1) \Gamma^{a}{}_{p(\uc}
                        \Gamma^{p}{}_{bd)}
\]
where $\uc$ contains $n>0$ indices. There will be cases where the number of hidden indices
needs to be made clear. In such cases one of two options can be taken -- either state
explicitly the number of hidden indices in words (as in the above example) or by including
the number as a subscript as in this example
\[
\Gamma^{a}{}_{b\uc_n d} = \Gamma^{a}{}_{(b\uc_n,d)}
                  - (n+1) \Gamma^{a}{}_{p(\uc_n}
                          \Gamma^{p}{}_{bd)}
\]
This notation can be extended to handle expressions such as
\[
\Gamma^a{}_{bc,i_1i_2i_3\cdots i_n}A^{i_1}A^{i_2}A^{i_3}\cdots A^{i_n}
\]
which would be written in condensed form as
\[
\Gamma^a{}_{bc,\ud}A^{\Xdot\ud}
\]
Note carefully the dot to the left of the $\ud$ superscript. Its purpose is to avoid an
ambiguity that would arise in using the condensed notation for an expression such as
\[
\Gamma^a{}_{bc,i_1i_2i_3\cdots i_n}A^{i_1i_2i_3\cdots i_n}
\]
which, following the earlier discussion, would be written as
\[
\Gamma^a{}_{bc,\ud}A^{\ud}
\]
These simple changes bring some degree of normalcy to the printed form but those gains
rapidly pale into insignificance when displaying the fifth order expansions generated by
Cadabra (in section (\ref{sec:fifthOrderRNC})). Cadabra's output does not employ the
familiar use of brackets to denote symmetrisation over an index list. Instead, it uses the
fully expanded form which, on paper, can lead to dramatic explosion in otherwise similar
looking terms. There seems little point in printing terms that differ only by swapping pairs
of indices. Thus the convention used here will be that symmetrisation over an index list on
the right hand side of an equation will be inferred from any explicit use of brackets for
symmetrisation written on the left hand side. Thus an equation like
\[
A_{(ab)} = B_a C_b
\]
will be taken to mean
\[
A_{(ab)} = \frac{1}{2!}\left (B_a C_b + B_b C_a\right)
\]
This convention will only be applied to results generated by Cadabra. The left hand side
will only contain a single term with just one pair of brackets.

The final notational device concerns cases where an index needs to be excluded from
symmetrisation. The normal practise is to enclose the index in a pair of vertical lines.
This involves two characters whereas a single dot over the chosen index can serve the same
purpose. Thus $(ab{\Dot c}d{\Dot e}fg)$ will denote symmetrisation over only $a,b,d,f$ and
$g$. In the standard notation this would have been written as $(ab\vert c\vert d\vert e\vert
fg)$.

\section{The metric in Riemann normal form}\label{sec:MetricRNC}

In the preceding section the generic and Riemann normal coordinates distinguished by using
the symbols $x^a$ and $y^a$ respectively. Now, for notational convenience and to accord with
convention, the Riemann normal coordinates will be denoted $x^a$ while $y^a$ will be
stripped of any special meaning.

The aim of this section is to express the metric in Riemann normal form. This will take the
form of an infinite series in powers of the curvature and its derivatives. Start by writing
out the Taylor series for the metric around $x^a=0$
\[
g_{ab}(x) = g_{ab} + \sum_{n=1}^\infty\> \frac{1}{n!}\>g_{ab,\uc}\>x^{\Xdot\uc}
\]
where $\uc$ contains $n$ indices and $g_{ab}$ are constants (e.g.
$g_{ab}=\diag(1,1,1,\cdots)$).

The present task is to express the partial derivatives of the metric in terms of the Riemann
tensor. From the standard definition of a metric compatible connection, a series of partial
derivatives leads to
\[
g_{ab,c\ud} = \left(g_{ae}\Gamma^{e}{}_{bc}+g_{eb}\Gamma^{e}{}_{ac}\right)_{,\ud}
\]
and since $g_{ab,c\ud}$ is totally symmetric in $c\ud$ it is easy to see that
\[
g_{ab,c\ud} = \left(g_{ae}\Gamma^{e}{}_{b(c}\right){}_{\!,\ud)}
            + \left(g_{eb}\Gamma^{e}{}_{a(c}\right){}_{\!,\ud)}
\]
Two points should be noted, first, the connection appears only in the form
$\Gamma^{a}{}_{b(c,\ud)}$, second, the left hand side contains derivatives one order higher
than in the corresponding terms on the right hand side. The upshot is that this equation can
be used to recursively compute all of the metric derivatives solely in terms of the
$\Gamma^{a}{}_{b(c,\ud)}$ and the constants $g_{ab}$. In this way the above Taylor series
for the metric can be expressed solely in terms of the connection and its derivatives. But
more can be done -- the derivatives of the connection must surely tie in with the
curvatures. Thus attention turns to the standard definition for the curvature, which after a
series of derivatives, can be written in the form
\begin{equation}
R^a{}_{(bc{\Dot d},\ue)} = \Gamma^a{}_{d(bc,\ue)} - \Gamma^a{}_{(bc,\ue)d}
                         + \left(\Gamma^a{}_{i(c}\Gamma^i{}_{b{\Dot d}}\right){}_{\!,\ue)}
                         - \left(\Gamma^a{}_{id}\Gamma^i{}_{(bc}\right){}_{\!,\ue)}
\pglabel{eqn:RabcdDerivs}
\end{equation}
(Note that the $\Gamma^a{}_{dbc}$ in $\Gamma^a{}_{d(bc,\ue)}$ is the first of the
generalised connections discussed earlier in section (\ref{sub:CoordsIVP})). Can this result
be used to eliminate the connection and its derivatives from the metric? Yes, but only after
specialising to the Riemann normal coordinates.

Recall that, in Riemann normal coordinates, all geodesics through $O$ are of the form
\[
x^a(s) = s v^a
\]
which upon substitution into the geodesic equations leads to
\[
0 = \Gamma^a{}_{(b\uc)}\qquad\text{at $O$}
\]
It follows, by recursion on equation \eqref{eqn:gamCovDeriv}, that
\begin{equation}
0 = \Gamma^a{}_{(bc,\ud)}\qquad\text{at $O$}
\pglabel{eqn:rncGamDeriv}
\end{equation}
Recall that $\Gamma^a{}_{bc,\ue d}$ is separately symmetric in its first pair of indices and
in the remaining $(n+1)$ lower indices (assuming $\ue$ contains $n$ indices). Thus using
equation \eqref{eqn:doubleSymm} it follows that
\[
0 = (n+3)\Gamma^a{}_{(bc,\ue d)} = 2\Gamma^{a}{}_{d(b,c\ue)}
                                 + (n+1)\Gamma^{a}{}_{(bc,\ue)d}
\]
This result can be used to eliminate the $\Gamma^a{}_{(bc,\ue)d}$ term in equation
\eqref{eqn:RabcdDerivs} for the derivatives of the curvature. The result, after a minor
shuffling of terms is
\[
(n+3)\Gamma^a{}_{d(b,c\ue)} =
        (n+1)\left( R^a{}_{(bc\Dot d,\ue)}
                   - \left(\Gamma^a{}_{i(c}\Gamma^i{}_{b{\Dot d}}\right){}_{\!,\ue)}
                   + \left(\Gamma^a{}_{id}\Gamma^i{}_{(bc}\right){}_{\!,\ue)}
             \right)
\]
(the reason for rearranging the terms will become clear in a moment). Note also that the
last term in the previous equation can be eliminated by equation \eqref{eqn:rncGamDeriv} and
the product rule.

In summary, the equations of interest are
\begin{gather}
\pglabel{eqn:metricTaylor}
g_{ab}(x) = g_{ab} + \sum_{n=1}^\infty\> \frac{1}{n!}\>g_{ab,\uc}\>x^{\Xdot\uc}\\[5pt]
\pglabel{eqn:metricGamma}
g_{ab,c\ud} = \left(g_{ae}\Gamma^{e}{}_{b(c}\right){}_{\!,\ud)}
             + \left(g_{eb}\Gamma^{e}{}_{a(c}\right){}_{\!,\ud)}\\[10pt]
\pglabel{eqn:GammaRiemann}
(n+3)\Gamma^a{}_{d(b,c\ue)} =
        (n+1)\left( R^a{}_{(bc\Dot d,\ue)}
                   - \left(\Gamma^a{}_{i(c}\Gamma^i{}_{b{\Dot d}}\right){}_{\!,\ue)}
              \right)
\end{gather}
These equations could be used as follows. First, equation \eqref{eqn:GammaRiemann} is used to
recursively compute the $\Gamma^a{}_{b(c,\ue d)}$ in terms of the Riemann tensor and its
partial derivatives (this was the reason behind the shuffling of terms noted above). Note
that $\ue$ in equation \eqref{eqn:GammaRiemann} contains $n$ hidden indices. The
$\Gamma^a{}_{b(c,\ud)}$ are then substituted into \eqref{eqn:metricGamma} which in turn is
used to recursively express all of the $g_{ab,\uc}$ in terms of the Riemann tensor and its
partial derivatives. When the dust settles the result is a finite series expansion for the
metric in terms of the Riemann tensor and its partial derivatives. The result, to 3rd-order
in the curvature, is
\begin{dmath*}g_{ab}(x) = \cdb{metric4} + \BigO{\eps^4}\end{dmath*}
Though this result meets the stated aim -- to express the metric in terms of the curvatures
and its derivatives -- there is a small problem. The partial derivatives make subsequent
raising and lowering of indices very tedious. A far better result would be one that employs
covariant derivatives. Fortunately the work required to achieve that is not too onerous. The
details are described in the remainder of this section.

It is not hard to see that a series of covariant derivatives of the Riemann tensor,
would lead to an equation of the form, in any coordinate frame,
\begin{equation*}
R^{a}{}_{(bc{\Dot d};\ue)} = R^{a}{}_{(bc{\Dot d},\ue)} + Q^{a}{}_{(bc{\Dot d}\ue)}
\end{equation*}
where $Q^{a}{}_{(bc{\Dot d}\ue)}$ is a function of the $\Gamma^{a}{}_{bc}$, the
$R^{a}{}_{bcd}$ and their partial derivatives. If this is going to sit nicely with the
algorithm given above then it will be necessary to show, in the Riemann normal frame, that
this equation only contains connection terms of the form $\Gamma^{p}{}_{q(r,\us)}$.
Fortunately this is rather easy to do.

Each term of the form $\Gamma^{p}{}_{qr,\us}$ in $Q$ arose during one round of covariant
differentiation. Thus \emph{at least one} of the indices $q,r$ and \emph{all} of the indices
in $\us$ must be drawn from the index list $\ue$. If both $q$ and $r$ are contained in $\ue$
then the term is of the form $\Gamma^{p}{}_{(qr,\us)}$ and thus will vanish when specialised
to the Riemann normal frame. This completes the proof. By re-arranging the above equation
into the following form
\begin{equation}
R^{a}{}_{(bc{\Dot d},\ue)} = R^{a}{}_{(bc{\Dot d};\ue)} - Q^{a}{}_{(bc{\Dot d}\ue)}
\pglabel{eqn:RabcdPartials}
\end{equation}
it can be used to recursively compute all of the partial derivatives of the curvatures in
terms of their covariant derivatives. The $Q^{a}{}_{(bc{\Dot d}\ue)}$ will contain lower
order derivatives of the curvatures and partial derivatives of the connection all of which
can be eliminated (in favour of covariant derivatives) using previously computed results.
For the first two derivatives it turns out that the partial and covariant derivatives are
equal (as expected) but differences do appear in higher order derivatives. These differences
will be apparent when the fifth order results are given later in section
(\ref{sub:partialRabcd}).

\section{The inverse metric in Riemann normal form}\label{sec:InvMetricRNC}

Most of the hard work is done and it is now time to develop algorithms for Riemann normal
expansions for other interesting quantities, in this instance the inverse metric
$g^{ab}(x)$. The previous section used $0=g_{ab;\uc}$ as a starting point to express the
metric in terms of the curvatures. On this occasion, for the inverse metric, the starting
point will be $0=g^{ab}{}_{;\uc}$. Then, following a path similar to that used in the
previous section, leads to the following equations
\begin{gather}
\pglabel{eqn:InvMetricTaylor}
g^{ab}(x) = g^{ab} + \sum_{n=1}^\infty\> \frac{1}{n!}\>g^{ab}{}_{,\uc}\>x^{\Xdot\uc}\\[5pt]
\pglabel{eqn:InvMetricGamma}
g^{ab}{}_{,c\ud} = - \left(g^{ae}\Gamma^{b}{}_{e(c}\right){}_{\!,\ud)}
                   - \left(g^{eb}\Gamma^{a}{}_{e(c}\right){}_{\!,\ud)}\\[10pt]
(n+3)\Gamma^a{}_{d(b,c\ue)} =
        (n+1)\left( R^a{}_{(bc\Dot d,\ue)}
                   - \left(\Gamma^a{}_{i(c}\Gamma^i{}_{b{\Dot d}}\right){}_{\!,\ue)}
              \right)\tag{\ref{eqn:GammaRiemann}}
\end{gather}
These equations can be used to construct the series expansion for $g^{ab}(x)$, which to
3rd-order is
\begin{dmath*}g^{ab}(x) = \cdb{metric4.inv} + \BigO{\eps^4}\end{dmath*}
%

\section{Generalised connections}\label{sec:GenGamma}

In section (\ref{sub:CoordsIVP}) the generalised connections $\Gamma^{a}{}_{bc\ud}$ were
shown to arise from successive differentiation of the geodesic equation and that they can be
computed recursively using
\[
\Gamma^{a}{}_{b\uc d} = \Gamma^{a}{}_{(b\uc,d)}
                - (n+1) \Gamma^{a}{}_{p(\uc}
                        \Gamma^{p}{}_{bd)}
\tag{\ref{eqn:GenGam}}
\]
where the index list $\uc$ contains $n>0$ indices.

Here are the first three generalised connections
%
\def\genGammaA{{\Gamma^{a}_{(bc)}(x)}}%
\def\genGammaB{{\Gamma^{a}_{(bcd)}(x)}}%
\def\genGammaC{{\Gamma^{a}_{(bcde)}(x)}}%
\def\genGammaD{{\Gamma^{a}_{(bcdef)}(x)}}%
%
\begin{dgroup*}[compact,spread={3pt}]
\begin{dmath*}\genGammaA = \cdb{genGamma04}+\BigO{\eps^4}\end{dmath*}%
\begin{dmath*}\genGammaB = \cdb{genGamma14}+\BigO{\eps^4}\end{dmath*}%
\begin{dmath*}\genGammaC = \BigO{\eps^4}\end{dmath*}
\end{dgroup*}
%

\section{Geodesics}\label{sec:GeodesicsRNC}

The discussion so far has been framed mostly around the set of geodesics that pass though
the point $O$. There are of course many geodesics within the patch of $O$ that do not pass
through $O$. The challenge now is to construct any geodesic within the patch of $O$. The
approach followed here will be similar to that taken previously in which solutions of the
geodesic initial value problem are used to solve the geodesic boundary value problem.

\subsection{Geodesic initial value problem}\label{sec:GeodesicIVP}

Consider a point $P$ distinct from $O$. At $P$ it is reasonable to assume that the
generalised connections do not vanish (which is generally true, the exception being flat
space). Thus the coordinates $x^a$ in the neighbourhood of $P$ do not constitute a Riemann
normal frame relative at $P$. But as $P$ lies in the patch for $O$ it follows (by definition
of the patch) that the metric is non-singular at $P$ and thus a new set of Riemann normal
coordinates, $y^a$, with $P$ as the origin, can always be constructed.

This problem has been discussed once before, in section (\ref{sub:CoordsBVP}). Using equation
\eqref{eqn:SolIVP} and the generalised connections from section (\ref{sec:GenGamma}) leads to
\begin{dmath*}x^{a}(s)
   = x^{a}
   + s \dotx^{a}
   + \frac{1}{24} s^2 {\Dot x}^b {\Dot x}^c \Bigl(\cdb{ivp42}\Bigr)
   + \frac{1}{12} s^3 {\Dot x}^b {\Dot x}^c {\Dot x}^d \bigl(\cdb{ivp43}\bigr)
   + \BigO{s^4,\eps^4}\end{dmath*}
%
\subsection{Geodesic boundary value problem}\label{sec:GeodesicBVP}

Consider now the case of three distinct points $O$, $P$ and $Q$. The goal in this section is
to compute the geodesic that passes through $P$ and $Q$. Start by using equation
\eqref{eqn:iterBVP} for the generalised connections from section (\ref{sec:GenGamma}) to
obtain
\begin{dgroup*}[compact,spread={3pt}]
\begin{dmath*} x^{a}(\lambda) = \cdb{bvp4} + \BigO{\eps^4} \end{dmath*}
\end{dgroup*}
where $\lambda=s/L_{PQ}$ is the scaled geodesic distance from $P(\lambda=0)$ to
$Q(\lambda=1)$.

\section{Geodesic arc-length}\label{sec:ArcLength}

Given the explicit expressions for the metric and the geodesic that joins the points $P$ and
$Q$, the length of that geodesic can be computed by way of the integral
\[
L_{PQ} = \int_P^Q\>\left(g_{ab}(x)\frac{dx^a}{ds}\frac{dx^b}{ds}\right)^{1/2}\>ds
\]
Up to this point the parameter $s$ has been taken to be the proper distance along the
geodesic. However, after careful inspection of the geodesic path \eqref{eqn:SolIVP} it is
clear that any uniform scaling of $s$ is allowed. Thus it is possible to re-scale $s$ so
that $s=0$ at $P$ and $s=1$ at $Q$ (of course, the parameter $s$ no longer measures proper
distance so it may be better to use a different symbol, say $\lambda$). Furthermore, a
standard result states that the integrand is constant along the geodesic and can thus be
evaluated at any point such as $P$. The integral is now trivial to evaluate with the result
\[
L^2_{PQ} = \left.g_{ab}(x)\frac{dx^a}{ds}\frac{dx^b}{ds}\right\vert_{P}
\]
which, using previous results, leads to
\begin{dmath*}L^2_{PQ} = \cdb{lsq4} + \BigO{\eps^4}\end{dmath*}
A simple calculation shows that this result can also be written in the slightly more
suggestive form
\[
L^2_{PQ} = g_{ab}(\bar x) Dx^a Dx^b + \BigO{\eps^4}
\]
where ${\bar x}^a = (x^a_P+x^a_Q)/2$ is the \emph{coordinate} mid-point of the geodesic.
This does raise the question -- Can higher order estimates for $L^2$ (see equation
\ref{eqn:geodesicLSQ}) be obtained by sampling the metric at suitablly chosen points on
(or near) the geodesic? That question will be left for another occassion.

\section{Cadabra}\label{sec:Cadabra}

Though the above equations seem simple they do impose a significant computational cost --
their recursive structure quickly leads to expressions that are too hard to manage (for
humans) beyond the first few iterations. The computations are best left to computer programs
specifically designed for tenor computations. One such program is Cadabra -- a {\tt C++}
program that reads plain text files to perform the various tensor computations. It was
initially designed for computations in high energy physics but it also very well suited to
tensor computations in general.

Cadabra's syntax is a hybrid of LaTeX to express tensor expressions, Python to coordinate the
computations and some unique Cadabra syntax to describe properties of various objects (e.g.,
index sets, symmetries, commutation rules etc.). A number of simple examples, including a
detailed discussion of Cadabra's syntax with particular emphasis for use in General
Relativity, can be found on the GitHub repository \cite{brewin:2019-01}.
This includes full Cadabra sources for all the examples and exercises (with solutions)
covering elementary topics (e.g., verifying $\nabla g=0$ for a metric connection) through
to more advanced topics (e.g., deriving the BSSN equations from the ADM equations).
This material was based on an earlier paper \cite{brewin:2009-02} (written for an
ealier version of Cadabra).

For those readers seeking a flavour of what Cadabra code looks like (before rushing off to
learn Cadabra in detail) here are two short examples. The first shows how Cadabra can be
used to verify that $\nabla g=0$ when $\nabla$ is the metric compatible covariant
derivative. The second example shows how the truncation operator $T^m_{\eps}$ introduced in
section (\ref{sub:CoordsBVP}) can be implemented in Cadabra.

\subsection{The metric connection}\label{sub:MetricGamma}

This simple example (adapted from the earlier paper \cite{brewin:2009-02})
verifies that $\nabla g =0$ when $\nabla$ is the metric compatible covariant derivative.

\lstset{numbers=left,gobble=2}
\CdbSetup{action=verbatim}

\begin{cadabra}
   # Define some properties

   {a,b,c,d,e,f,h,i,j,k,l,m,n,o,p,q,r,s,t,u#}::Indices.

   g_{a b}::Metric.
   g_{a}^{b}::KroneckerDelta.

   \partial_{#}::PartialDerivative.

   # Define a rule for the Christoffel symbol

   Gamma := \Gamma^{a}_{b c} -> (1/2) g^{a d} (  \partial_{b}{g_{d c}}
                                               + \partial_{c}{g_{b d}}
                                               - \partial_{d}{g_{b c}} );

   # Define the covariant derivative of the metric

   cderiv := \partial_{c}{g_{a b}} - g_{a d}\Gamma^{d}_{b c}
                                   - g_{d b}\Gamma^{d}_{a c};

   # Do the computations

   substitute          (cderiv, Gamma);
   distribute          (cderiv);
   eliminate_metric    (cderiv);
   eliminate_kronecker (cderiv);
   canonicalise        (cderiv);
\end{cadabra}

The code is rather easy to follow. The first few lines defines a set of indices (line 3) and
some objects with particular properties (lines 5 to 8). This is followed by two rules for
constructing a connection (line 12) and the covariant derivative of $g_{ab}$ (line 18). The
body of the calculations can be seen in the last few lines (23 to 27) where various
operations (known as \emph{algorithms} in Cadabra's lexicon) are applied. As expected, the
above code returns a value of zero for the object \cdbverb{cderiv}.

A few points are worth noting. Cadabra uses a hybrid syntax of Python and LaTeX. Tensor
equations are defined and results are returned using a subset of LaTeX while computations
(such as substitutions, simplifications etc.) are expressed using a Python syntax. This use
of Python and LaTeX makes for an easy entry into Cadabra programming as most users would
already have familiarity with Python and LaTeX. Using LaTeX to define and record
tensors means that it is very easy to carry results from one Cadabra notebook to another or
even to other LaTeX documents -- all of the Cadabra output shown in this paper were imported
without change from the Cadabra output generated by other documents.

\subsection{Truncation of polynomials}\label{sub:Truncation}

In section (\ref{sub:CoordsBVP}) a truncation operator $T^m_{\eps}$ was introduced. The
question here is -- How might that operator be implemented in Cadabra?

Suppose you are asked to extract the leading terms from an expression such as
\[
P^a(x) =  c^a
        + c^a_{b} x^b
        + c^a_{bc} x^b x^c
        + c^a_{bcd} x^b x^c x^d
        + c^a_{bcde} x^b x^c x^d x^e
\]
One approach (there are others, e.g., emulating a truncated Taylor series) is to use
Cadabra's \verb!::Weight! property and the \verb!keep_weight! algorithm. The idea is to
assign weights to nominated objects (through the \verb!::Weight! property) and then extract
terms matching a chosen weight (using the \verb!keep_weight! algorithm).

Here is a small piece of Cadabra code that does the job.
\begin{cadabra}
   def truncate (obj,n):

       # define the weight and give it a label
       x^{a}::Weight(label=\epsilon).

       # start with an empty espression
       ans = Ex(0)

       # loop over selected terms in the source
       for i in range (0,n+1):

          foo := @(obj).
          bah  = Ex("\epsilon = " + str(i))

          # extract a single term
          keep_weight (foo, bah)

          # update the running sum
          ans = ans + foo

       # all done, return final answer
       return ans

   # the quartic polynomial
   quarticPoly :=   c^{a}
                  + c^{a}_{b} x^{b}
                  + c^{a}_{b c} x^{b} x^{c}
                  + c^{a}_{b c d} x^{b} x^{c} x^{d}
                  + c^{a}_{b c d e} x^{b} x^{c} x^{d} x^{e}.

   # truncate to cubic terms
   cubicPoly = truncate (quarticPoly,3)
\end{cadabra}
The first thing to note is the Python function \cdbverb{truncate}. This simply does as its
name suggests -- it truncates an object to a certain order. How does it do the job? The
first line to note is line 4. This identifies $x^a$ as the target to carry the weights (and
is given the label \verb!\epsilon! to distinguish it from other targets declared by other
instances of \verb!::Weight!). Cadabra now sees the polynomial $P^a(x)$ as if it had been
written as
\[
P^a(x) =   c^a
         + c^a_{b} x^b \eps
         + c^a_{bc} x^b x^c \eps^2
         + c^a_{bcd} x^b x^c x^d \eps^3
         + c^a_{bcde} x^b x^c x^d x^e \eps^4
\]
The Python \cdbverb{for loop} then extracts the requested terms from the source returning
the final truncated answer in \cdbverb{ans}. The final result is exactly as expected -- the
leading cubic part of the original quartic polynomial. This function and variations on it
(e.g., extracting a single term) are used extensively in all of the Cadabra codes used in
this paper.

\section{Expansions to fifth order}\label{sec:fifthOrderRNC}

All of the $\BigO{\eps^6}$ Cadabra programs were not overly demanding on computational
resources, taking between 0.5 second and 12 minutes to run and requiring
between 40 and 360 Mbyte of memory (on a Mac Pro (2013) running macOS 10.14.4).

The Cadabra codes and several support scripts are available from the author's GitHub site
\ \url{https://github.com/leo-brewin/riemann-normal-coords}.

Some of the following expansions can be compared directly with results obtained by
traditional methods. In particular the expansions for the metric and its inverse, equations
\eqref{eqn:metric} and \eqref{eqn:metricinv} agree exactly with those given in equations
(17) and (18) respectively of \cite{hatzinikitas:2000-01}. Also, equation
\eqref{eqn:genGamma0} agrees with that given in equation (A12) of \cite{calzetta:1988-01}
(though showing that they are equivalent does require some work using the first and second
Bianchi identities).

Note that in the following output any symmetrisation of indices on the right hand side
should be inferred from the explicit symmetrisation (the round brackets) on the left hand
side (as per the discussion in section \ref{sec:Notation}).

\clearpage


\def\mysubsection#1{\vskip1.5\parskip\leftline{\bf #1}\vskip-1.5\parskip\ignorespaces}
\def\dMath#1#2#3#4{\begin{dmath*}[number={#1}]#2\pglabel{#3}#4(#1)\end{dmath*}}

\mysubsection{The metric}

\begin{dgroup*}[compact,spread={3pt}]
   \dMath{11.1}{g_{(a b)}(x) = \cdb{metric}+\BigO{\eps^6}}{eqn:metric}{\hfill}
\end{dgroup*}

\mysubsection{The inverse metric}

\begin{dgroup*}[compact,spread={3pt}]
   \dMath{11.2}{g^{(a b)}(x) = \cdb{metric.inv}+\BigO{\eps^6}}{eqn:metricinv}{\hfill}
\end{dgroup*}

\mysubsection{The metric determinant}

\begin{dgroup*}[compact,spread={3pt}]
   \dMath{11.3a}{-g(x)         = \cdb{Ndetg}     + \BigO{\eps^6}}{eqn:detg}{\hskip 1cm\hfill}
   \dMath{11.3b}{\sqrt{-g(x)}  = \cdb{sqrtNdetg} + \BigO{\eps^6}}{eqn:sqrtdetg}{\hskip 1cm\hfill}
   \dMath{11.3c}{\log{(-g(x))} = \cdb{logNdetg}  + \BigO{\eps^6}}{eqn:logdetg}{\hfill}
\end{dgroup*}

\clearpage

\mysubsection{Generalised connections}

\vskip 10pt

The following results were obtained by recursive application of equation \eqref{eqn:GenGam}.
\begin{dgroup*}[compact,spread={3pt}]
   \dMath{11.4}{\Gamma^{a}_{(b c)}(x) = \cdb{genGamma0} + \BigO{\eps^6}}{eqn:genGamma0}{\hfill}
   \dMath{11.5}{\Gamma^{a}_{(b c d)}(x) = \cdb{genGamma1} + \BigO{\eps^6}}{eqn:genGamma1}{\hfill}
   \dMath{11.6}{\Gamma^{a}_{(b c d e)}(x) = \cdb{genGamma2} + \BigO{\eps^6}}{eqn:genGamma2}{\hfill}
   \dMath{11.7}{\Gamma^{a}_{(b c d e f)}(x) = \cdb{genGamma3} + \BigO{\eps^6}}{eqn:genGamma3}{\hfill}
\end{dgroup*}

\mysubsection{Symmetrised partial derivatives of the connection.}\label{sub:partialGamma}

\vskip 10pt

The following results were obtained from equation \eqref{eqn:GammaRiemann}
\begin{dgroup*}[compact,spread={3pt}]
   \dMath{11.8}{   3 \Gamma^a{}_{d(b,c)} = \cdb{dGamma61}}{eqn:dGamma61}{\hfill}
   \dMath{11.9}{   6 \Gamma^a{}_{d(b,ce)} =  \cdb{dGamma62}}{eqn:dGamma62}{\hfill}
   \dMath{11.10}{ 15 \Gamma^a{}_{d(b,cef)} =  \cdb{dGamma63}}{eqn:dGamma63}{\hfill}
   \dMath{11.11}{  9 \Gamma^a{}_{d(b,cefg)} = \cdb{dGamma64}}{eqn:dGamma64}{\hfill}
   \dMath{11.12}{252 \Gamma^a{}_{d(b,cefgh)} = \cdb{dGamma65}}{eqn:dGamma65}{\hfill}
\end{dgroup*}

\mysubsection{Symmetrised partial derivatives of the %
              Riemann curvature tensor.}\label{sub:partialRabcd}

\vskip 5pt

\vskip 10pt

The following results were obtained from equation \eqref{eqn:RabcdPartials}. Note that by
inspection of these results and those above it is easy to verify the claim made in
section (\ref{sec:MetricRNC}) that the connection terms that arise in
$R^{a}{}_{(bc{\Dot d};\ue)}$ are always of the form $\Gamma^{p}{}_{q(r,\us)}$.
\begin{dgroup*}[compact,spread={3pt}]
   \dMath{11.13}{    R^a{}_{(cd{\Dot b},e)} = \cdb{dRabcd61}}{eqn:dRabcd61}{\hfill}
   \dMath{11.14}{    R^a{}_{(cd{\Dot b},ef)} = \cdb{dRabcd62}}{eqn:dRabcd62}{\hfill}
   \dMath{11.15}{ -2 R^a{}_{(cd{\Dot b},efg)} = \cdb{dRabcd63}}{eqn:dRabcd63}{\hfill}
   \dMath{11.16}{ -5 R^a{}_{(cd{\Dot b},efgh)} = \cdb{dRabcd64}}{eqn:dRabcd64}{\hfill}
   \dMath{11.17}{ -3 R^a{}_{(cd{\Dot b},efghi)} = \cdb{dRabcd65}}{eqn:dRabcd65}{\hfill}
\end{dgroup*}

\mysubsection{Riemann normal coordinates}

\begin{equation*}
   y^a = \ny{0}^{a} + \ny{1}^{a} + \ny{2}^{a} + \ny{3}^{a} + \ny{4}^{a}\tag{11.18}
\end{equation*}

\begin{dgroup*}
   \dMath{11.18a}{    \ny{0}^{a} = \cdb{rnc61}}{eqn:rnc61}{\hfill}
   \dMath{11.18b}{  2 \ny{1}^{a} = \cdb{rnc62}}{eqn:rnc62}{\hfill}
   \dMath{11.18c}{  6 \ny{2}^{a} = \cdb{rnc63}}{eqn:rnc63}{\hfill}
   \dMath{11.18d}{ 24 \ny{3}^{a} = \cdb{rnc64}}{eqn:rnc64}{\hfill}
   \dMath{11.18e}{360 \ny{4}^{a} = \cdb{rnc65}}{eqn:rnc65}{\hskip 1cm}
\end{dgroup*}

\mysubsection{Geodesic IVP}

\begin{align}
   x^{a}(s) ={}& x^{a}
            + s {{\Dot x}^a}
            + \frac{s^2}{2!} {{\Dot x}^b} {{\Dot x}^c} A^{a}_{bc}
            + \frac{s^3}{3!} {{\Dot x}^b} {{\Dot x}^c} {{\Dot x}^d} A^{a}_{bcd}
            + \frac{s^4}{4!} {{\Dot x}^b} {{\Dot x}^c} {{\Dot x}^d}
                                                       {{\Dot x}^e} A^{a}_{bcde}\notag\\
           &+ \frac{s^5}{5!} {{\Dot x}^b} {{\Dot x}^c} {{\Dot x}^d}
                                                       {{\Dot x}^e}
                                                       {{\Dot x}^f} A^{a}_{bcdef}
            + \BigO{s^6,\eps^6}\tag{11.19}
\end{align}
\begin{dgroup*}
   \dMath{11.19a}{ 360 A^{a}_{bc} = \cdb{ivp62}}{eqn:ivp62}{\hfill}
   \dMath{11.19b}{ 360 A^{a}_{bcd} = \cdb{ivp63}}{eqn:ivp63}{\hskip 0.5cm}
   \dMath{11.19c}{  90 A^{a}_{bcde} = \cdb{ivp64}}{eqn:ivp64}{\hfill}
   \dMath{11.19d}{   3 A^{a}_{bcdef} = \cdb{ivp65}}{eqn:ivp65}{\hfill}
\end{dgroup*}

\mysubsection{Geodesic BVP}

\begin{align}
   x^{a}(s) &= x^{a} + s Dx^{a}
                     + (s-s^2) x^{a}_2
                     + (s-s^3) x^{a}_3
                     + (s-s^4) x^{a}_4
                     + (s-s^5) x^{a}_5
                     + \BigO{s^6,\eps^6}\tag{11.20}\pglabel{eqn:bvp6}\\
   x^{a}_2 &= \nx{2}^{a}_2 + \nx{3}^{a}_2 + \nx{4}^{a}_2 + \nx{5}^{a}_2
            + \BigO{\eps^6}\tag{11.20a}\\
   x^{a}_3 &= \nx{3}^{a}_3 + \nx{4}^{a}_3 + \nx{5}^{a}_3 + \BigO{\eps^6}\tag{11.20b}\\
   x^{a}_4 &= \nx{4}^{a}_4 + \nx{5}^{a}_4 + \BigO{\eps^6}\tag{11.20c}\\
   x^{a}_5 &= \nx{5}^{a}_5 + \BigO{\eps^6}\tag{11.20d}
\end{align}
\begin{dgroup*}[spread=5pt]
   \dMath{11.20a.1}{    -3 \nx{2}^{a}_2 = \cdb{bvp622}}{eqn:bvp622}{\hfill}
   \dMath{11.20a.2}{   -24 \nx{3}^{a}_2 = \cdb{bvp623}}{eqn:bvp623}{\hfill}
   \dMath{11.20a.3}{  -720 \nx{4}^{a}_2 = \cdb{bvp624}}{eqn:bvp624}{\hskip 1cm}
   \dMath{11.20a.4}{  -360 \nx{5}^{a}_2 = \cdb{bvp625}}{eqn:bvp625}{\hfill}
   \dMath{11.20b.1}{   -12 \nx{3}^{a}_3 = \cdb{bvp633}}{eqn:bvp633}{\hfill}
   \dMath{11.20b.2}{  -720 \nx{4}^{a}_3 = \cdb{bvp634}}{eqn:bvp634}{\hskip 1cm}
   \dMath{11.20b.3}{ -1080 \nx{5}^{a}_3 = \cdb{bvp635}}{eqn:bvp635}{\hskip 1cm}
   \dMath{11.20c.1}{  -180 \nx{4}^{a}_4 = \cdb{bvp644}}{eqn:bvp644}{\hfill}
   \dMath{11.20c.2}{ -2160 \nx{5}^{a}_4 = \cdb{bvp645}}{eqn:bvp645}{\hskip 1cm}
   \dMath{11.20d.1}{  -360 \nx{5}^{a}_5 = \cdb{bvp655}}{eqn:bvp655}{\hskip 1cm}
\end{dgroup*}

\mysubsection{Geodesic arc-length}

\begin{equation*}
   \pglabel{eqn:geodesicLSQ}
   \left(\Delta s\right)^2 = \nD{0} + \nD{2} + \nD{3} + \nD{4} + \nD{5}
                           + \BigO{\eps^6}\tag{11.21}
\end{equation*}

\begin{dgroup*}[spread=5pt]
   \dMath{11.21a}{      \nD{0} = \cdb{lsq60}}{eqn:lsq60}{\hfill}
   \dMath{11.21b}{    3 \nD{2} = \cdb{lsq62}}{eqn:lsq62}{\hfill}
   \dMath{11.21c}{   12 \nD{3} = \cdb{lsq63}}{eqn:lsq63}{\hfill}
   \dMath{11.21d}{  360 \nD{4} = \cdb{lsq64}}{eqn:lsq64}{\hfill}
   \dMath{11.21e}{ 1080 \nD{5} = \cdb{lsq65}}{eqn:lsq65}{\hfill}
\end{dgroup*}

\section{Source}

All of the Cadabra files used in preparing this paper are available at the GitHub repository
\ \url{https://github.com/leo-brewin/riemann-normal-coords}

\section{Acknowledgements}

I am very grateful to Kasper Peeters for his many helpful suggestions. Any errors,
omissions or inaccuracies in regard to Cadabra are entirely my fault.

\appendix

\section{Symmetrisation of tensors}

The totally symmetric part of a tensor $A_{i_1i_2i_3\cdots i_n}$ is commonly
defined by
\[
A_{\left(i_1i_2i_3\cdots i_n\right)}
= \frac{1}{n!}\left(  A_{i_1i_2i_3\cdots i_n}
                    + A_{i_1i_2i_3\cdots i_n}
                    + A_{i_1i_2i_3\cdots i_n}
                    + \cdots \right)
\]
where the sum on the right hand side includes every permutation of the indices
of $i_1i_2i_3\cdots i_n$. If the tensor $A_{i_1i_2i_3\cdots i_n}$ happens to be
symmetric in every pair of indices then
\[
A_{\left(i_1i_2i_3\cdots i_n\right)} = A_{i_1i_2i_3\cdots i_n}
\]
From the above definition it is very easy to establish the following theorems
\begin{gather*}
A_{(i_1i_2i_3\cdots(j_1j_2j_3\cdots j_m)\cdots i_n)}
= A_{(i_1i_2i_3\cdots j_1j_2j_3\cdots j_m\cdots i_n)}\\[5pt]
A_{(i_1i_2i_3\cdots i_n}B_{j_1j_2j_3\cdots j_m)}
= A_{((i_1i_2i_3\cdots i_n)}B_{(j_1j_2j_3\cdots j_m))}\\[5pt]
nA_{(i_1i_2i_3i_4\cdots i_n)} = A_{i_1(i_2i_3i_4\cdots i_n)}
                        + A_{i_2(i_1i_3i_4\cdots i_n)}
                        + A_{i_3(i_1i_2i_4\cdots i_n)}
                        + \cdots\notag\\[5pt]
                        {\hskip 0.7\textwidth} + A_{i_n(i_1i_2i_3\cdots i_{n-1})}\\[5pt]
nA_{(i_1i_2i_3i_4\cdots i_n)} = A_{(i_2i_3i_4\cdots i_n)i_1}
                        + A_{(i_1i_3i_4\cdots i_n)i_2}
                        + A_{(i_1i_2i_4\cdots i_n)i_3}
                        + \cdots\notag\\[5pt]
                        {\hskip 0.7\textwidth} + A_{(i_1i_2i_3\cdots i_{n-1})i_n}
\end{gather*}

Suppose now that $A_{i_1i_2i_3\cdots i_n} = A_{(i_1i_2i_3\cdots i_n)}$, that is,
$A_{i_1i_2i_3\cdots i_n}$ is totally symmetric. Then for any $B_j$
\begin{gather*}
\begin{split}
(n+1) A_{(i_1i_2i_3\cdots i_n}B_{j)}
&= A_{ji_2i_3\cdots i_n} B_{i_1}
+ A_{i_1ji_3\cdots i_n} B_{i_2}
+ A_{i_1i_2j\cdots i_n} B_{i_3}\\[5pt]
&\quad+ \cdots + A_{i_1i_2i_3\cdots i_{n-1}j} B_{i_n}
\end{split}\\[5pt]
\intertext{and}
\begin{split}
(n+1) A_{(i_1i_2i_3\cdots i_n,j)}
&= A_{ji_2i_3\cdots i_n,i_1}
+ A_{i_1ji_3\cdots i_n,i_2}
+ A_{i_1i_2j\cdots i_n,i_3}\\[5pt]
&\quad+\cdots
+A_{i_1i_2i_3\cdots i_{n-1}j,i_n}
+A_{i_1i_2i_3\cdots i_n,j}
\end{split}
\end{gather*}

All of the above are very easy to prove but one result which requires just a little more
thought is the following.

Suppose $A_{i_1i_2j_3j_4j_5\cdots j_n}$ is symmetric in the pair $i_1i_2$ and symmetric in
all the indices $j_3j_4j_5\cdots j_n$. That is, it is symmetric under the interchange of any
pair of $i$'s and any pair of $j$'s but it is \emph{not} necessarily symmetric when any $i$
is swapped with any $j$. What can be said about $A_{(i_1i_2j_3j_4j_5\cdots j_n)}$? Here is
the result
\begin{gather}
n A_{(i_1i_2i_3\cdots i_n)}
= 2 A_{i_n(i_1i_2i_3\cdots i_{n-1})}
+ (n-2) A_{(i_1i_2i_3\cdots i_{n-1})i_n}
\pglabel{eqn:doubleSymm}
\end{gather}
The proof is very easy. Begin by writing out $n!A_{(i_1i_2i_3\cdots i_n)}$ in full. Then
partition the terms into two disjoint sets, one set in which $i_n$ appears in one of the
first two index slots, the other set in which $i_n$ appears in any of the remaining $n-2$
slots. The terms in the first set are exactly those that define $A_{i_n(i_1i_2i_3\cdots
i_{n-1})}$ while those in the second set define $A_{(i_1i_2i_3\cdots i_{n-1})i_n}$. The
above equation follows by simply counting the number of terms in each set ($2(n-1)!$ and
$(n-2)(n-1)!$ respectively) and the simple observation that $n!A_{(i_1i_2i_3\cdots i_n)}$
equals the sum of the terms from both sets.

Finally note that if $Q=A_{i_1i_2i_3\cdots i_n}x^{i_1}x^{i_2}x^{i_3}\cdots x^{i_n}$
then
\begin{gather*}
Q_{,i_1i_2i_3\cdots i_n} = n! A_{(i_1i_2i_3\cdots i_n)}\\[10pt]
Q = A_{(i_1i_2i_3\cdots i_n)}x^{i_1}x^{i_2}x^{i_3}\cdots x^{i_n}
\end{gather*}

\printbibliography  


\end{document}